\documentclass[prc,showpacs,preprint,nofootinbib]{revtex4}
\usepackage{amsmath}
\usepackage{epsfig}
\usepackage{color}

\newcommand{\re}{\ref}
\newcommand{\be}{\begin{equation}}
\newcommand{\ee}{\end{equation}}

\newcommand{\la}{\label}
\newcommand{\ber}{\begin{eqnarray}}
\newcommand{\eer}{\end{eqnarray}}

\begin{document}

\title{The Lorentz Integral Transform and its Inversion}
\author{N. Barnea$^{a}$, V.D. Efros$^{b}$, W. Leidemann$^{c}$, and G. Orlandini$^{c}$}

\affiliation{
$^{a}$Racah Institute of Physics, Hebrew University, 91904, Jerusalem,
Israel\\
$^{b}$Russian Research Centre "Kurchatov Institute", 123182 Moscow,  Russia\\
$^{c}$Dipartimento di Fisica, Universit\`{a} di Trento and INFN\\
(Gruppo Collegato di Trento), via Sommarive 14, I-38100 Trento, Italy\\
}
\date{\today}


\begin{abstract}
The Lorentz integral transform method is briefly reviewed.
The issue of the inversion of the transform, and in particular its ill-posedness, is addressed. 
It is pointed out that the mathematical term {\it ill-posed} is misleading and merely due
to a historical misconception. In this connection standard regularization procedures for the solution
of the integral transform problem are presented. In particular 
a recent one is considered in detail and critical comments on it are provided.
In addition a general remark concerning the concept of the Lorentz integral transform as a method with a {\it controlled resolution} is made. 
\end{abstract}


\maketitle

\section{Introduction}

The Lorentz integral transform (LIT) method \cite{ELO94} allows the calculation of observables
for reactions into the many-body continuum without an explicit use of a many-body continuum
wave function. The LIT approach has had a wide application range in the field of  
electroweak reactions with few-nucleon systems. In fact calculations of 
inclusive inelastic reactions of nuclei with 
$A=3$ (see e.g. \cite{3bgam,3bee'1,3bee'2,3bee'3}), $A=4$ (see e.g. \cite{4bgam1,4bgam2,4bgamSonia,4bgamsofia,4bnu,4bee'1,4bee'2}) and $A=6,7$ \cite{6bgam,7bgam}
have been carried out. In some cases also exclusive reactions  have been considered \cite{4bexgam,4bexee'1,4bexee'2}. Various successful comparisons
to results of conventional calculations, with explicit continuum state wave functions, 
have been performed for the two- and three-body systems \cite{ELO94,3bee'1,Lapiana,Golak,ELOB07,ELOT09}.
A recent review of the LIT method is given in \cite{ELOB07}.

To illustrate the LIT approach we consider the following inelastic response function
\begin{equation}\label{response}
R(\omega)=\sum\!\!\!\!~\!\!\!\!\!\!\!\!\int _f\,\,
|\langle f|\Theta| 0 \rangle |^2 \delta(E_f-E_0-\omega) \, .
\end{equation}
Here $E_0$ and $|0\rangle$ are ground state energy and wave function of the
particle system under consideration, $E_{f}$ and $|f\rangle$ denote final state energy
and wave function of the final particle system, and $\Theta$
is the operator inducing the response function $R$.
The LIT of the response $R$ is given by
\begin{equation} \label{LIT}
{\cal L}(\sigma_R,\sigma_I) = 
\int d\omega {\frac {R(\omega)} {(\omega -\sigma_R)^2 + \sigma_I^2}} \,.
\end{equation}
The key point of the method is that ${\cal L}$ can be calculated without any knowledge
of the final state wave functions $|f\rangle$. In fact the transform is given by
\begin{equation}
{\cal L}(\sigma_R,\sigma_I) = \langle \tilde\Psi | \tilde\Psi \rangle 
\end{equation}
and the Lorentz function $\tilde\Psi$ is obtained from
\begin{equation} \label{LITeq}
(H-E_0-\sigma_R - i \sigma_I)| \tilde \Psi \rangle = \Theta |0\rangle \,.
\end{equation}
The solution of Eq.~(\ref{LITeq}) is unique and has an asymptotic boundary condition similar to a bound state
and thus can be solved with bound-state methods.
As final step, the response $R(\omega)$ is obtained by inverting Eq.~(\ref{LIT}).
Such an inversion, however, represents a so called {\it ill-posed} problem. 
For non-experts in the field of ill-posed problems the question may arise if the ill-posedness does
not lead to a serious drawback of the LIT approach.
Therefore the aim of this paper is a clarification of the mathematical term {\it ill-posed}
and to outline how ill-posed problems can be reliably solved.

\section{Ill-posed problems}

To illustrate the general situation regarding the ill-posed problems, let us first consider an example that shows how ill-posed problems 
are quite common. This is the  problem of calculating the derivative $g(x)$ of a function $f(x)$ in accordance with the definition  
\be g(x)={\cal D}f(x);\qquad {\cal D}f(x)={\rm lim}_{\delta\rightarrow0}\frac{f(x+\delta)-f(x-\delta)}{2\delta}.\la{der}\ee  
This very familiar problem is an ill-posed one, since it does not satisfy a requirement  characteristic 
for a {\it well-posed problem}  as defined by Hadamard in 1902. In fact the function $g(x)$ does not  depend continuously 
on the input $f(x)$. For example, consider  $F(x)= f(x)+\Delta f(x)$, then $G(x)= g(x)+\Delta g(x)$. 
If  $\Delta f(x)=\epsilon\sin(\nu x)$, it is evident that for any small
value of $\epsilon$, $\nu$ values can be found that make $\Delta g(x)={\cal D}\Delta f(x)$ large.

This means that the problem of the derivative of any function calculated numerically is ill-posed, since 
this function inevitably  includes a random noise. 

What is done in this case is to replace Eq.~(\re{der}) 
with its analogue,  but with a finite $\delta$ (denoted by $ d$):
\be G(x,d)={\hat D(d)}f(x);\qquad {\hat D(d)}f(x)=\frac{f(x+d)-f(x-d)}{2d}.\la{de1}\ee 
The quantity $G(x,d)$ is called a regularized solution to the problem, and $d$ is called a regularization parameter. 
In contrast to $g(x)$ from Eq.~(\re{der}), $G(x,d)$ depends continuously on $f(x)$.
Thus, for sufficiently small increments $\Delta f(x)$ the increments $\Delta g ={\hat D(d)}\Delta f$ will also be small. At the same time, one can
find  $d$ values such that the right--hand side of Eq.~(\re{de1}) with the true, noiseless, $f(x)$ differs from the true derivative $g(x)={\cal D} f(x)$ 
by an arbitrarily small amount. As a result, if one considers a sequence of noise increments $\Delta f(x)$ with amplitudes  
tending to zero one can choose a sequence of 
corresponding $d$ parameters
in a way that the arising sequence of regularized solutions $G(x,d)$ will tend to the true solution $g(x)$. 
 
It is easy to obtain that, at a given noise increment $\Delta f(x)$, $d$ values  of  about $[2\Delta f(x)/f''(x)]^{1/2}$ in $G(x,d)$
provide an optimal approximation to
the true derivative ${\cal D} f(x)$. At such $d$ values,
the error in ${\cal D} f(x)$ is about $[2f''(x)\Delta f(x)]^{1/2}$.
When $\Delta f(x)$ tends to zero this error tends to zero as well.   
From what has been  said above one can infer that 
the $d$ dependence of $G(x,d)$ at a given $x$ value should include a plateau at an appropriate level of  accuracy, 
and any value of $d$ belonging to this plateau can be 
adopted for representing ${\cal D} f(x)$.

The situation is quite similar when one deals with the problem of finding a solution to the integral equation of the first kind (Eq.~(\ref{LIT}) is a specific example)
\be L(y)=\int_0^\infty K(y,x)R(x)dx, \la{ie1}\ee 
or, in short, $L={\hat K}R$.
The case when there exists a unique solution is considered. 
The role of the operator ${\cal D}$ from Eq.~(\re{der}) 
is played here by the operator ${\hat K}^{-1}$. 
We also assume that either a) $\int_0^\infty |K(y,x)|dx$ is convergent or
b) $\int_0^a |K(y,x)|dx$ is convergent at any finite $a$
and $K(y,x)\rightarrow0$ monotonically at $x\rightarrow\infty$.
The solution $R$ does not depend continuously on the input $L$. Indeed, let us replace $L$ with $L+\Delta L$ and 
suppose that an increment $\Delta L(y)$ is such that it causes a given 
increment $\Delta R=A\sin(kx)$  to the solution $R$. When
$|k|$ is sufficiently large the required $\Delta L$ is arbitrarily small, see e.g. \cite{BO}. 
Therefore,
Eq.~(\re{ie1}) poses an ill-posed problem \footnote{This consideration implies use of the metrics $C$ in the space of $R(x)$ and the
metrics $C$ or the metrics $L_2$ in the space of $L(y)$.}.  
One may note that this property is a
very general one so that a special explanation of the ill-posedness of the  transform when the kernel is 
a Lorentzian function is superfluous.

Similar to the example of the derivative mentioned above, Eq.~(\re{ie1}) can be solved by a regularization method.
The regularization suppresses the unphysical high--frequency components that might arise in an approximate solution.
The problem (\re{ie1}) is replaced with a well--posed
problem that includes a {\it regularization parameter} ``$r$''. The solution $R(x,r)$ of this  {\it well--posed} problem depends on $L(y)$ continuously and thus 
sufficiently small
increments in $L(y)$ lead to only small increments in $R(x,r)$. At the same time, with the true $L(y)$,  the solution $R(x,r)$ is arbitrarily
close to
$R(x)$ at a proper choice of $r$. 
If a sequence of inputs $L(y)$ with increasing accuracy is considered then choosing $r$ in a proper way one has a sequence of
solutions $R(x,r)$ tending in the limit to the true $R(x)$. The corresponding $r$ values may be found using estimates of magnitudes of uncertainties
 in $L(y)$
and estimates on how close   $R(x,r)$ is to $R(x)$ at a given $r$. This  can also be done without such estimates, relying on 
stability of $R(x,r)$ when $r$ varies.

For a number of regularization methods the mentioned convergence of $R(x,r)$ to $R(x)$ is proven, and in spite of the unfortunate term {\it ill posed}, 
which sounds negative the approach is {\it well--founded} (see e.g. 
\cite{TA77,LRS86,TGSY95}).  The origin of this term 
lies merely in the history and has by no means a relation to the accuracy of the results obtained by the outlined approach. 
The term was introduced about hundred years ago by the mathematician Hadamard  who suggested that such problems cannot have a physical relevance. 
As we know nowadays, this was definitely a mistake as it is claimed by modern authors. For example at p. 2 of \cite{LRS86}, the authors 
write "It turned out that the opinion
of Hadamard regarding the Cauchy problem for the Laplace equation and a number of other problems of the same type was erroneous."

\section{Solutions of ill-posed problems: the regularization procedures}

Among the number of regularization schemes for solving Eq.~(\re{ie2}) that are known, the one that has been 
applied in conjunction with few--body calculations of the Lorentz transform allows reliable estimates 
of uncertainties of the results and  keeps the uncertainties at a sufficiently low level. 
In all those applications the solution has been represented as the sum of the first $N$ functions  of a complete set $\{\phi_i(x)\}$,
\be R(x,N)=\sum_{i=1}^Nc_i\phi_i(x).\la{er}\ee 
Here the value $N$ plays the role of the regularization parameter. The coefficients $c_i$ are found from fitting $L(y)$
(for more detail see \cite{ELOB07}). As pointed out in the introduction 
accurate solutions have been obtained using
this procedure. On the one hand the accuracy is shown by the existence of a range of $N$ values where the solution is stable, and on the other hand
by the already mentioned various benchmark calculations. 

 In particular we would like to mention that with this regularization procedure
 it is very easy to implement the small $x$ 
behaviour of $R(x)$, allowing the inclusion of the Coulomb case and the amplification of the role of the threshold region in the fitting procedure. 
The discussed regularization approach of Eq.~(\ref{er}) is well suited to describe
responses $R(x)$ with a single--peak structure. Different methods might be
advantageous in other cases (see also \cite{Andreasi}).

One may also note that in this regularization scheme input 
values of $L(y)$ from a limited range of $y$ is sufficient. 
This range is comparable to the range of $x$ values of interest in $R(x)$. 
In addition one can easily  work with different LIT resolution parameters 
$\sigma_I$ in different $y$ ranges (see e.g. \cite{ELOT_highq}). 

In conclusion, the regularization procedure used in few--body calculations allows reliable estimates 
of uncertainties of the results as well as keeping the uncertainties at a sufficiently low level. 

In this context it is interesting that in a recent paper the claim is made 
that the inversion problem of any ill-posed integral transform of type
\be L(y)=\int_0^\infty K(y-x)R(x)dx.\la{ie2}\ee
can be tackled in a conceptually completely new 
manner, namely without any use of regularization techniques~\cite{GS09}.
Since from a conceptual point of view this would be a very remarkable result, leading to an
alternative method to invert the LIT, we will consider the proposed method in the following.

First of all we note that as a matter of fact the ``novel inversion method'' in~\cite{GS09}
is in principle known, see e.g. \cite{TA77}. It goes as follows.
Eq.~(\re{ie2}) is solved  via application of the Fourier transformation. This gives 
\begin{equation}
 {\tilde R}(k)=(2\pi)^{-1/2}{\tilde L}(k)/{\tilde K}(k) \,,
\end{equation}
where the Fourier transform of a function $f(x)$ is denoted by $\tilde f(k)$.
In this way, the problem turns to the problem of numerical inversion of the Fourier transform ${\tilde R}(k)$. When calculated numerically this includes
an admixture of errors whose relative magnitude
increases at high frequencies. This happens because at large $|k|$ values ${\tilde K}\rightarrow0$ and the errors in the input ${\tilde L}$ are amplified. 
Therefore the contribution of large $|k|$ to the integrand when inverting the Fourier transform 
should be suppressed in some way. 

We note in passing that the criterion of {\it well--posedness} mentioned in~\cite{GS09} for the case of 
kernels which are functions of a difference, like those in Eq.~({\ref{ie2}), can never be satisfied. In fact the authors
claim that only when $ inf |{\tilde K}(k)|>0$, $\tilde K$ being the Fourier
transform of the kernel, the problem is well-posed. However that condition can never be fulfilled for kernels 
that satisfy the natural conditions for existence of their Fourier transform \cite{BO},
i.e. either a) $\int_{-\infty}^\infty|K(s)|ds$ is convergent or 
b) such an integral over any finite interval is convergent and $K(s)\rightarrow0$ monotonically at $s\rightarrow\pm\infty$. 
Indeed,  when these conditions are satisfied one 
always has ${\tilde K}(k)\rightarrow0$ at $k\rightarrow\pm\infty$. 
 
In order to diminish the above mentioned faulty contribution to the inversion of ${\tilde R}(k)$
at large $|k|$, in~\cite{TA77}
it is recommended to insert a damping factor $f(k,k_0)$ in the integrand  
where $k_0$ is a regularization parameter. The simplest possibility
is to 
cut the integrand at some $|k|= k_{\rm max}$ getting, see e.g. \cite{VS},
\be R(x,k_{\rm max})=\frac{1}{2\pi}\int_{-k_{\rm max}}^{k_{\rm max}}\frac{{\tilde L}(k)}{{\tilde K}(k)}e^{-ikx}dx.\la{rap}\ee

In~\cite{GS09} a modification of this procedure is used, where in addition 
${\tilde R}(k)$ with $|k|>k_{\rm max}$ is taken into account approximately. 
This is achieved by representing ${\tilde R}(k)$ in the high $|k|$ region with its leading 
asymptotics matched to the above ${\tilde R}(k)$  at $k=\pm k_{\rm max}$. The leading 
asymptotics are obtained from the known small $x$ asymptotics of $R(x)$ 
(if $R(x)\sim x^\lambda$ at $x\rightarrow0$ 
then ${\tilde R}(k)\sim |k|^{-(1+\lambda)}$ at $|k|\rightarrow\infty$, see e.g. \cite{AE}).
In fact, in general the small $x$ behaviour of a physical relevant $R$ is 
known, and is, as already pointed out before, always implemented in the inversions of the above mentioned few-body applications. 
The criterion in \cite{GS09} to choose the $k_{\rm max}$ value could probably be improved
by considering the stability of $R(x,k_{\rm max})$ with respect to variations in
$k_{\rm max}$. 

If the error--admixture amplitude in ${\tilde R}(k)$ is sufficiently small then 
the contribution of this admixture to the approximate solution (\re{rap})
or to such a quantity with the asymptotics contributions added, is also small. At the same time, in the absence of an admixture of errors,
 the corresponding approximate solution $R(x,k_{\rm max})$  of the form (\re{rap})
tends to the true $R(x)$ when $k_{\rm max}$ tends to infinity.  
Thus, this is nothing else but a usual regularization procedure. 
Just as such, it is presented
 e.g. in \cite{TA77,VS}. The said is, of course, true irrespective
to whether one adds the approximate leading asymptotics contribution, or not.      
 
The inversion method put forward in \cite{GS09} can be be added to the list
of already existing inversion techniques (see \cite{Andreasi}).
However, the method has not been applied in \cite{GS09} to a realistic test case, therefore
one does not know its quality. In particular it might be problematic to rely on the assumption that the threshold
behaviour of the response function dominates the high $|k|$ behaviour of
$\tilde R(k)$ obtained numerically. In fact if the high $|k|$ region is dominated by numerical errors
then the matching procedure will fail. Such an undesirable behaviour has been found
applying just this  inversion technique, for example, to the $^4$He photodisintegration LIT \cite{LB09}. 
In this realistic case the range of $x$ values dominated by the asymptotics turns out to be too narrow. 
Consequently, the true $\tilde R(k)$ falls off quite fast
and the $|k|$ values where its asymptotics start are higher.  
  
Another possible drawback of this method could be the fact that in order to calculate $\tilde L(k)$
the input $L(y)$ is required in a very wide range of $y$. 
This is a disadvantage since the precision of the calculation of the LIT $L(y)$
is correlated with $y$. In particular at  a fixed relative precision it is much more expensive to obtain $L(y)$ for large $y$.

Finally, we would like to make an additional remark concerning the general philosophy 
of the LIT also in relation to its ill-posed character. The LIT
method has to be understood as an approach with a {\it controlled resolution}. If
one expects that $R$ has structures of a width $\Delta$ then the LIT resolution
parameter $\sigma_I$ should be similar in size. Then it is sufficient to determine the corresponding LIT 
with a moderately high precision. If in reality  no structures with a width smaller than $\Delta$
are present the inversion will lead to reliable results for $R(x)$. If, however, there is a reason to believe that $R(x)$ exhibits
such smaller structures one should
reduce $\sigma_I$ accordingly and perform again a calculation of $L(y)$ with
the same relative precision as before. Such a calculation is of course more expensive than
the previous one with larger $\sigma_I$, but in principle one can reduce
the LIT resolution parameter $\sigma_I$ more and more.
It is worth noticing that LIT procedure is similar to what is done in a conventional
continuum calculation. In this case one calculates $R(x)$ for a finite number of $x$ points which are then
connected, e.g., by a spline interpolation.  In doing so, also here one assumes that there is no structure between two neighbouring grid points.  
If, however, there is reason to believe that $R(x)$ exhibits such structures the density of the grid points 
has to be increased (normally easier in a conventional calculation).

The advantage
of the LIT approach as compared with a conventional approach is evident. In the
LIT case the finite resolution makes the calculations feasible, also when a 
conventional calculation which corresponds to an infinite resolution ($\sigma_I \rightarrow 0$) is not
feasible any more.

\end{document}